\documentclass[12pt]{iopart}
\usepackage{graphicx}
\newcommand{\Fig}[1]{Fig.~\ref{#1}}

\begin{document}

\title{Electron recombination with tungsten ions with open f-shells}
\author{C Harabati$^1$, J C Berengut$^1$, V V Flambaum$^{1,2}$, and V A Dzuba$^{1}$}
\address{$^1$ School of Physics, University of New South Wales,
Sydney 2052, Australia}
\address{$^2$ Helmholtz Institute Mainz, Johannes Gutenberg University, 55099 Mainz, Germany}
\ead{c.harabati@unsw.edu.au, v.flambaum@unsw.edu.au}


\begin{abstract}
We calculate the electron recombination rates with target ions  W$^{q+}$, $q = 18$ -- $25$, as functions of electron energy and electron temperature (i.e. the rates integrated over the Maxwellian velocity distribution). Comparison with available experimental data for W$^{18+}$, W$^{19+}$, and W$^{20+}$ is used as a test of our calculations. Our predictions for  W$^{21+}$, W$^{22+}$, W$^{23+}$, W$^{24+}$, and W$^{25+}$ (where the experimental data are not available) may be used for plasma modelling in thermonuclear reactors. The results for the temperature dependent rates for each ion are fitted with the standard analytical expressions to make them easy to use. All of these ions have an open electron $f$-shell and have an extremely dense spectrum of chaotic many-electron compound resonances which enhance the recombination rates by 2-3 orders of magnitude in comparison with the direct electron recombination. Conventional dielectronic recombination theory is not directly applicable in this case. Instead, we developed a statistical theory based on the properties of chaotic eigenstates. This theory describes a multi-electronic recombination (extension of the dielectronic recombination) via many-excited-electron compound resonances.   
\end{abstract}
\maketitle
%
%
\section{Introduction} 

The aim of this work is to calculate an electron recombination rate with tungsten ions that are impurities in the fusion plasma. The current theoretical database for the rate coefficients from ADAS (Atomic Data and Analysis Structure \cite{ADAS, BOSA11}) is not reliable.
Tungsten is a vital element for ITER (International Thermonuclear Experimental Reactor) and JET(Joint European Torus) since it is used as plasma-facing material and for a divertor of the machine, which collects the heat produced by fusion plasma \cite{PCEHK13, RJC15}. Thus some contamination of the core plasma by tungsten ions is unavoidable. Detailed knowledge of the reaction rates of tungsten ions with electrons in plasma (such as excitation, ionisation, and recombination) is very important for the fusion plasma modelling and an efficient energy production from the fusion plasma.   

The problem with multiply charged tungsten ions and many other ions with several electrons in an open $f$-shell is that there is a very dense spectrum of chaotic multi-electronic compound resonances there which dominates the recombination rate. The interval between these resonances may be smaller than 10$^{-6}$~eV, and the total number of important configuration state functions (CSFs) may reach billions \cite{BHDFG15}. In these circumstances, their mixing becomes chaotic and extremely sensitive to any unaccounted perturbation (such as higher order correlation and quantum electrodynamics corrections)~\cite{FGGK94}.

Traditional approaches to the calculation based on the dielectronic recombination are not directly applicable here. Instead, we must deal with the multi-electronic recombination. We have developed a consistent statistical theory --- the many-body quantum chaos (MBQC) statistical theory~\cite{FGGK94,GGF99,DFGH12} --- in which the electron capture cross section into the multi-electronic resonances is expressed in terms of the dielectronic ``doorway'' resonances broadened by the coupling to the multi-electronic excitations. The recombination cross section is obtained as a product of this capture cross-section and the fluorescent yield for the multi-electronic compound states \cite{FKG15}. The MBQC statistical theory predicts the results averaged over a small energy interval containing many compound resonances. Due to the very dense spectrum of the compound resonances and finite experimental energy resolution, this averaging also happens in the experimental data.

Previously, we studied electron recombination in Au$^{25+}$ using the statistical theory~\cite{GGF99,FGGH02}. 
These works found that the measured $\sim100$-fold enhancement of the recombination rate over the predicted direct radiative rate at low energies (less than or about 1 eV) \cite{HUSF98} is explained by the dense spectrum of compound resonances containing chaotic mixtures of many electron basis states. A similar enhancement was calculated for the recombination with W$^{20+}$~\cite{DFGH12} that is also in agreement with measurements at the heavy-ion storage ring~\cite{SBM11}. In~\cite{DFGHK13} we calculated the fluorescence yield in the same framework which enabled us to calculate the electron recombination rate over a larger electron energy range. Finally, our most recent work in the series examined the electron capture cross section in the many-electron determinant basis and in the basis of configuration state functions with definite total electron angular momentum~\cite{BHDFG15}. The result shows that both methods give nearly the same outcome for the electron capture via compound resonance states. 

In this paper we apply our MBQC statistical theory to the problem of electron recombination with multiply charged tungsten ions with open $f$-shells. In Sections~\ref{sec:spectrum} and \ref{sec:statistical} we summarize the theory, including specific details of relevance for these ions.  
Electron recombination rates with W$^{20+}$ and nearby ions W$^{18+}$ and W$^{19+}$ have been measured and calculated in Refs.~\cite{SBM11, BBGO12, SBKN14,BSKN16, S15}. Comparison of our calculations with the experimental data from these papers is presented in \ref{sec:results}. 
We also predict the electron recombination rates for the other ions W$^{q+}$ ($q=21$ -- 25) where experimental data are not available and future measurements may be prohibitively complicated or even impossible using present methods. These ions may have significant abundance in the plasma at certain temperatures (about $100-300$ eV for W$^{20+}$ \cite{SBM11}). Finally we convolve the recombination cross sections with an isotropic Maxwellian electron energy distribution to obtain thermally-averaged recombination rates for all eight ions for use in plasma modelling. These are presented in Figures~\ref{fig:temrate1} and \ref{fig:temrate2} and are the main result of this work. The temperature -dependent recombination rate at temperatures about 100 eV is not sensitive to the exact positions of the doorway resonances since the integration over energy smooths out all structures which have ~ 10 eV scale . 

Atomic units ($\hbar = |e| = m_e = 1$) are used throughout this work except where otherwise specified.

\section{Comparison of di-electronic and chaotic multi-electronic recombination}
For many years the di-electronic recombination theory was used to calculate the recombination rate. Therefore, it is important to explain why this approach should not be applied for ions with many active  electrons in open shells  where the processes are dominated by a very dense spectrum of compound resonances. Indeed, the capture of electron by a highly charged ion produces an excited state $\sim$1000 eV above the ground state. This energy may be distributed among  several open shell electrons in millions of different ways, and the di-electronic  states with only two excited electrons comprise a very small fraction of all available states. The number of available many-excited-electron states near the ionisation threshold is especially large when the $f$-shell is open. It is important that in the regime of a very strong chaotic configuration mixing all compound resonances can capture electron with comparable probabilities and  contribute to the resonance cross sections.

  However, even in this regime of chaotic compound resonances the di-electronic components of the compound states play a special role: they provide doorways to the compound states since they are directly coupled to the initial target+ electron state by the Coulomb interaction matrix element. Using the optical theorem we demonstrated \cite{FKG15} that the total resonance capture cross section of electron  is given by the sum over di-electronic doorways states only (see Eq. (\ref{eq:ccs})). However, these di-electronic resonances  are strongly broadened by the interaction with many-electron excitations, and the Breit-Wigner resonance denominator contains an additional spreading width $\Gamma_{\rm spr}$ (defined in Eq. (\ref{eq:sw})) which is significantly larger then the autoionisation and radiation widths.
  
  A similar conclusion is valid for other reactions, e.g. for the resonance photon capture (including the photoionisation) where the doorway state is the single-electron excitation produced by the direct photon capture from the atom ground state.
  
  The recombination cross section is obtained as a product of this capture cross-section and the fluorescent yield for the multi-electronic compound states. Here the difference between the di-electronic and multi-electronic recombination theories is even more significant. The fluorescent yield for the chaotic compound states is significantly higher than that for the di-electronic states. Indeed, for  a di-electronic state the numbers of the electron emission channels and the photon emission channels are comparable. On the contrary, in a chaotic compound state only a small fraction of the compound states components, the di-electronic components,  give the autoionzation width $\Gamma^{(a)}$ while all the compound state components contribute to the radiative width $\Gamma^{(r)}$ making the fluorescent yield (the ratio  $\Gamma^{(r)}/( \Gamma^{(r)} +\Gamma^{(a)}$)) strongly enhanced. In other words, the enhancement factor is the number of the radiative decay channels (given by  a million of the final states below the excited compound states) to a few open autonisation channels available slightly above the ionisation threshold.
   
\section{Spectrum and wavefunctions}
\label{sec:spectrum}

In this work we have applied the statistical theory to the recombination of an electron with open $f$-shell tungsten ions. The target ions are W$^{q+}\ (\textrm{[Kr]}4d^{10}4f^{n})$ with $q=$18 -- 25, and $n=28 - q$ (i.e. $n$ runs from 10 -- 3 for the respective ions). After the target and a free electron in continuum are recombined, the compound ions W$^{(q-1)+}\ (\textrm{[Kr]}4d^{10}4f^{n+1})$ are formed.

We began by solving the Dirac-Hartree-Fock (DHF) equations self-consistently for the core of the target ion $W^{q+}$. In this frozen $1s^2 ... 4f^n$ core we calculated a basis of single-particle orbitals  up to $7spdfg$. That is, we include all spectroscopic orbitals with principal quantum number $n \leq 7$ and orbital angular momentum $l \leq 4$.
This is sufficient for the calculation of the excitation spectrum near the ionization threshold of the compound ion W$^{(q-1)+}$.

The second step is to find a list of many-electron configurations for target and compound ions using this basis by distributing the electrons of the $4d$ and $4f$ shells among the 31 relativistic ($j=l \pm 1/2$) orbitals ($5s$ -- $7s$, $5p$ -- $7p$, $4d$ -- $7d$, $5f$ -- $7f$, and $5g$ -- $7g$). There are $(10+n)$ electrons and $(11+n)$ electrons for the target and compound ions, respectively. 
Each configuration has an average energy $E^\textsl{CA}$ and represents $N$ many-electron states with various angular momenta $J$ and projections. Configuration mixing does not significantly affect the position of the configuration average energy.
Therefore, we may estimate the spectrum density of the excited states in the compound ion $W^{(q-1)+}$ by means of the average configuration energies $E^\textsl{CA}$ and  the number of many-electron states within each configuration $N$:
\begin{equation}
\label{eq:CA_energy}
E^\textsl{CA}= E_\textrm{core} + \sum_\alpha \epsilon_\alpha n_\alpha + \sum_{\alpha \leq \beta} \frac{n_\alpha (n_\beta - \delta_{\alpha\beta})}{1 + \delta_{\alpha\beta}} U_{\alpha\beta}
\end{equation}
\begin{equation}
\label{eq:N}
N = \prod_\alpha \frac{[j_\alpha]!}{n_\alpha! ([j_\alpha] - n_\alpha)!} \,,
\end{equation}
where the relativistic orbital $\alpha$ has the integer occupation number $n_\alpha$, $\left[j_\alpha\right]=2j_\alpha+1$ is the degeneracy of the orbital $\alpha$, and $\epsilon_\alpha$ is the single-particle energy of the orbital $\alpha$ in the DHF field of the core.
$U_{\alpha\beta}$ is the  average Coulomb interaction energy (including the exchange energy)  for electrons in states $\alpha$ and $\beta$:
\begin{equation}
\label{eq:Uab}
U_{\alpha\beta} = \frac{[j_\alpha]}{[j_\alpha]-\delta_{\alpha\beta}} \Bigg[ R_0 (\alpha\beta\alpha\beta) \\
	-\sum_\lambda \xi(l_\alpha + l_\beta + \lambda) R_\lambda (\alpha\beta\beta\alpha)
	\left(\begin{array}{ccc} j_\alpha            &  j_\beta          &\lambda\\
	                                    \frac{1}{2} &-\frac{1}{2} & 0 
	        \end{array} \right)^2 \Bigg].
\end{equation}
Here $R_\lambda (\alpha\beta\gamma\delta)$ is the radial integral of the two-body Coulomb interaction with a multipolarity $\lambda$~(\Eref{eq:radial_integral}), and   $\xi (L)=[1+(-1)^L]/2$ is the parity factor.

\begin{figure}[t!]
\centering
\includegraphics*[width=0.9\textwidth]{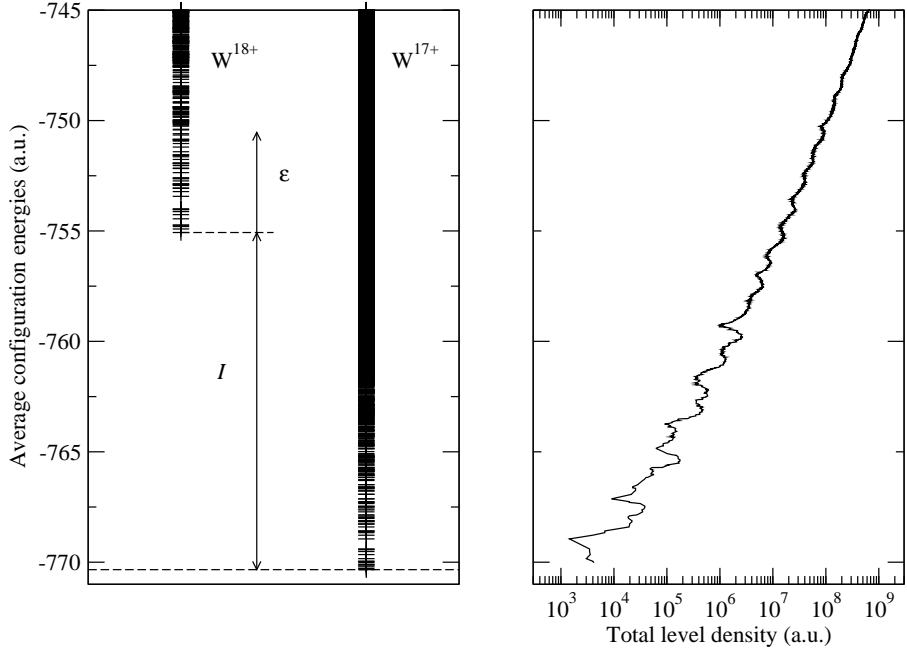}
\caption{Left panel: calculated configuration-average energy spectra of the target ion W$^{18+}$ (left) and the compound ion W$^{17+}$ (right). The ionisation energy $I$ and the electron ion collisional energy $\varepsilon$ have been marked on the graph. Right panel: the total level density of the compound ion W$^{17+}$ on the logarithmic scale.}
\label{fig:spectrum}
\end{figure}

Eigenvalues of the Dirac-Coulomb Hamiltonian operator in the many-electron basis can generally be found by diagonalising the matrix $H_{ij}=\langle\varphi_i|H|\varphi_j\rangle$ in the many-electron configuration basis $|\varphi_n\rangle$ (e.g. Slater determinants). The eigenvectors are expressed via expansion coefficients $C_i^{(\nu)}$ as a linear superposition $|\nu\rangle = \sum_i C_i^{(\nu)} |\varphi_i\rangle.$ This is the usual configuration interaction (CI) approach.
However, performing a configuration interaction calculation can be a formidable task for an atomic system which has many active valence electrons. This is the regime where a strong configuration mixing may happen due to the two-body residual electron-electron interaction. The level density of excited many-body states gets high and the residual interaction between electrons exceeds the mean level spacing  between the many-body states with fixed angular momentum $J$ and parity $\pi$. This is the reason for the emergence of the many-body quantum chaos \cite{BM69,FGGK94,ZBFH96}.

Fortunately, the configuration mixing can be taken into account statistically in the MBQC formalism. This approach has a very important advantage in that it does not require diagonalization of the huge Hamiltonian matrices \cite{F93, FV93, FG95, FG00}. The expansion coefficients for a complex compound state $|\nu\rangle$  have a random distribution with zero mean, $\overline{C_n^{(\nu)}}=0$. 
The variance $\overline{|C_n^{(\nu)}|^2}$  is given by the Breit-Wigner formula
\begin{equation}
\label{eq:breit-wigner}
\overline{|C_n^{(\nu)}|^2}=\frac{D}{2\pi}\frac{\Gamma _{\rm spr}}{(E_n-E_\nu )^2 +\Gamma _{\rm spr}^2/4},
\end{equation}
where $E_n=H_{nn}$ is the energy of a many-electron basis state, $E_\nu$ is the energy of a compound state, and $D$ is the mean level spacing between adjacent basis states (or compound resonances) in a given $J^\pi$ manifold. The spreading width $\Gamma_{\rm spr}$ can be expressed in terms of the non-diagonal matrix elements of the Hamiltonian matrix  \cite{BM69}: 
\begin{equation}
\label{eq:sw}
\Gamma_{\rm spr}= 2 \pi\frac{ \overline{|H_{nk}|^2}}{D}.
\end{equation}
The distribution \Eref{eq:breit-wigner} is normalised:  $\sum_n |C_n^{(\nu)}|^2\simeq\int\overline{|C_n^{(\nu)}|^2}dE_n/D=1$. The principal components are located within the energy interval $|E_n-E_\nu | \simeq \Gamma_{\rm spr}$, and the tail of the distribution for $|E_n-E_\nu| \gg \Gamma_{\rm spr}$ decreases in accordance with perturbation theory:  $|C_n^{(\nu)}|^2 \sim D \Gamma_{\rm spr}/(E_n-E_\nu )^2$. The emergence of quantum chaos and the distribution of the components \eref{eq:breit-wigner} have previously been tested in numerical calculations \cite{FGGK94,GGF99,GS03}. 

The left-hand panel of \Fig{fig:spectrum} shows the configuration-average energy spectra of the target ion W$^{18+}$ and the compound state W$^{17+}$. The average energy of each configuration and the number of many-electron states associated with it have been calculated from Eqs.~\eref{eq:CA_energy} and \eref{eq:N}, and these are used to calculate the total level density of the compound ion W$^{17+}$, shown in the right-hand panel of \Fig{fig:spectrum}.
The total level density of the compound ion near the ionisation threshold can be estimated by counting the number of many-electron states within the spreading width $\Gamma_{\rm spr}$, $\rho(E)\approx \sum_i N_i(E_i)/\Gamma_{\rm spr}$, where $E-\Gamma_{\rm spr}/2<E_i<E+\Gamma_{\rm spr}/2$.

In \Tref{t:W} our calculated ionisation energies are compared with NIST data~\cite{NIST}. Our calculations use the configuration-averaged approach; for example, the configuration-average ground state energy of W$^{18+}(4f^{10}) $ is $-755.0708$ a.u. and after recombination, the ground configuration energy of W$^{17+}(4f^{11})$ is $-770.3405$ a.u. Thus the ionisation energy is 15.27 a.u. in this approximation.

\Table{\label{t:W}Properties of tungsten ions. Here $I$ is the ionisation energy of the compound ion (a.u.), $N_{{\rm W}^{q+}}$ is the number of relativistic configurations within an energy range of 240 eV above the ionisation threshold of W$^{(q-1)+}$, $N_{{\rm W}^{(q-1)+}}$ is the number of configurations used to obtain density of states and total radiative width of the compound  ion. The spreading widths $\Gamma_{\rm spr}$ (a.u.) for each ion are also presented in the last column.} 
\br
\br
Target ion &\centre{2}{$I_{{\rm W}^{(q-1)+}}$} & $N_{{\rm W}^{q+}}$ & $N_{{\rm W}^{(q-1)+}}$ & $\Gamma^{\rm c}_{\rm spr}$ \\
\ns
&\crule{2}&&&\\
W$^{q+}(4f^n)$ & Expt.$^{\rm a}$& Theory$^{\rm b}$ & & &  \\
\mr 
W$^{18+}(4f^{10})~ ^5I_8$          & 15.5 & 15.27 & 227 & 92417  & 0.56 \\
W$^{19+}( 4f^9)~ ^6H^o_{15/2}$ & 17.0 & 16.73 & 165 & 30900  & 0.65 \\
W$^{20+}(4f^8)~ ^7F_6$              & 18.5 & 18.25 & 121& 52669  & 0.68 \\
W$^{21+}(4f^7)~ ^8S^o_{7/2}$    & 20.0 & 19.81  & 80  & 17782  & 0.68 \\
W$^{22+}(4f^6)~ ^7F_0$             & 21.8 &  21.41 & 59  & 13603   & 0.65 \\
W$^{23+}(4f^5)~ ^6H^o_{5/2}$    & 23.5 & 23.36  & 49  & 10617  & 0.59 \\
W$^{24+}(4f^4)~ ^5I_4$               & 25.2 & 25.05  & 35  & 8031    & 0.50 \\
W$^{25+}(4f^3)~ ^4I^o_{9/2}$      & 27.0 & 26.79  & 25  & 5769    & 0.16 \\ 
\br
\end{tabular}
\item[] $^{\rm a}$ Ionization energy of the compound ion from Ref.~\cite{NIST}.
\item[] $^{\rm b}$ Ionization energy in the mean field approximation used in this work.
\item[] $^{\rm c}$ The spreading widths have been calculated in Ref.~\cite{DFGH12}.
\end{indented}
\end{table}

We have found 227 configurations of  W$^{18+}(4f^{10}) $ in the energy range up to  $\varepsilon=240$ eV. These configurations provide final states for the autoionization channels which contribute to the total autoionization widths of the compound states. On the other hand, the energy range of the configurations making the compound states is located much higher, near the ionisation threshold. They are used to find the total level density and the total radiative width \eref{eq:Gamma_r}. The total number of configurations used for the compound ions are presented in the fifth column of  \Tref{t:W}.
 
The last column in Table \ref{t:W} shows the estimated spreading widths from Ref.~\cite{DFGH12}. It characterises the strength of the configuration mixing (see \Eref{eq:breit-wigner}). It can be understood as an energy range around a resonance from where the most of the doorway states' contributions comes. Outside the spreading width the coefficients $C_n^{(\nu)}$ decrease significantly. The spreading width  is expressed via mean-squared off-diagonal elements of the Hamiltonian matrix from~\eref{eq:sw}.
When more configurations are added to  the Hamiltonian matrix, both $\overline{H^2_{ij}}$ and the mean level spacing $D$ decrease. However their ratio does not change significantly,  so the spreading width remains  unchanged (for a given ion).

\section{Statistical theory applied to electron recombination}
\label{sec:statistical}

In previous papers it has been shown how the MBQC statistical theory can be used to calculate transition matrix elements between chaotic compound states \cite{F93,FV93,FG95,FG00} and to obtain the resonant cross-section of electron-ion recombination \cite{GGF99, FGGH02, DFGH12, DFGHK13,FKG15, BHDFG15}. We very briefly  present the main results  in this section. 

The theoretical resonant recombination process \cite{MB42} may be depicted as follows:   
\begin{equation}
\label{eq:recombination}
 e^-+A^{q+}  \rightarrow A^{(q-1)+**}\rightarrow A^{(q-1)+*}+\gamma .
\end{equation}
The first step of the process shows that the projectile electron is captured in a resonance state of the compound ion. The system is stabilized radiatively in the second step. The first step is reversible (by the autoionization).

In the standard resonance theory (see, e.g., Ref. \cite{LL77}), the recombination cross section is a fraction of the total (capture) cross-section: $\sigma=\omega_f\sigma_c$. The factor $\omega_f$ is called the branching ratio or the fluorescence yield since the autoionization and the radiative decay compete with each other in the process (\ref{eq:recombination}). This factorization enables one to deal with the calculations of the capture cross-section and the fluorescence yield separately. Note that values of $\omega_f$ may be different for different resonances, however in the statistical theory we actually calculate an energy-dependent average value of $\omega_f$.

\subsection{Capture cross-section}


Using the optical theorem expressing  the total cross section via the forward scattering amplitude and \eref{eq:breit-wigner} for the compound state wave functions, we have shown in Ref.~\cite{FKG15} that the summation over compound resonances may be replaced by the sum over dielectronic doorway states (the basis components which are directly connected to the continuum by the Coulomb interaction). The resonant capture cross-section in the statistical theory is given by
\begin{equation}
\label{eq:ccs}
\sigma_c= \frac{2\pi^2}{k^2}\sum_n\frac{(2J_n+1)}{2(2J_i+1)}
\frac{1}{2\pi}\frac{\Gamma _{\rm spr} \Gamma^{(a)}_{n i}}{(\varepsilon -\varepsilon _n)^2 +\Gamma _{\rm spr}^2/4}, 
\end{equation}
where $k$ is the wave number ($k^2=2\varepsilon$ in a.u.) of the projectile electron with collision energy $\varepsilon$, $\varepsilon_n=E_n-E_i$ is the energy of the dielectronic resonance relative to the ionisation threshold $E_i$ of the ion $A^{(q-1)+}$ 
( i.e. relative to the ground state energy of the target ion $A^{q+}$)
 and the autoionization width of the dielectronic resonance  $\varphi_n$ to the initial channel is given by
\begin{equation}
\label{eq:Gamma_ni}
\Gamma^{(a)}_{n i}
 = 2\pi \sum_{jl} |\langle (\varepsilon_n jl; J_i)J_n M_n | V |\varphi_n\rangle|^2 \,,
\end{equation} 
where only partial waves that satisfy the selection rules for the Coulomb interaction $V$ can contribute to the width. Here the continuum state of the free electron with the angular momentum $j$ and the target state with the angular momentum $J_i$ are coupled to construct the state with the total angular momentum and parity equal to that of the resonance $\varphi_n$. 

Note that according to \eref{eq:ccs} the width of the dielectronic resonance is dominated by the spreading width $\Gamma _{\rm spr}$ which characterises an `internal decay': i.e. the mixing of the dielecronic state with other several-excited-electron components of the compound state. Indeed, the spreading width $\Gamma _{\rm spr} \sim 0.5-0.7$ a.u. in the tungsten ions \cite{DFGH12} significantly exceeds the autoionization and radiative widths. If necessary, one may replace $\Gamma _{\rm spr}$ by the total width of the doorway state $n$:
$\Gamma_n = \Gamma _{\rm spr} + \Gamma^{(a)}_{n } + \Gamma^{(r)}_{n }$,
which includes the total autoionization and radiative widths \cite{FKG15}.

In our previous work \cite{BHDFG15} we calculated the resonance capture cross section using \eref{eq:ccs} and the basis of the dielectronic doorway states with definite angular momentum $J_n$ and projection (configuration state functions). We also performed calculations in a simpler basis of the Slater determinants built from the products of the single particle orbitals (which are not angular momentum eigenstates). The numerical calculations for W$^{20+}$ have demonstrated that the results of both calculations are in a good agreement. Therefore, in the present work  we use the Slater determinants as the basis of the dielectronic doorway states. The total resonance capture cross section in this basis is given by 
\begin{eqnarray}
\label{eq:capture-explicit}
 \sigma _c  = \frac{\pi^2}{k^2}\sum _{\alpha\beta\gamma^{-1}}\sum_{jl}
\Biggl[ 
\sum _\lambda \frac{X_\lambda[(jl)\gamma\alpha\beta]^2}{2\lambda +1}
+\sum _{\lambda\lambda '}\left\{ {j_\gamma 
\atop j }{j_\beta \atop j_\alpha }{\lambda \atop \lambda '}\right\}
X_\lambda[(jl)\gamma\alpha\beta] X_{\lambda'}[(jl)\gamma\beta\alpha] \nonumber\\
+\alpha\leftrightarrow\beta \Biggr] 
\frac{n_\gamma }{[j_\gamma]}\left(1-\frac{n_\alpha }{[j_\alpha]}\right)\left(1-\frac{n_\beta}{[j_\beta]}\right)\frac{\Gamma _{\rm spr}}{(\varepsilon -\varepsilon _{\alpha\beta\gamma^{-1}} )^2 +\Gamma _{\rm spr}^2/4},
\end{eqnarray}
where $X_\lambda[\delta\gamma\alpha\beta]$ is the Coulomb matrix element

\begin{eqnarray}
\label{eq:redmael}
X_\lambda[c\gamma\alpha\beta]
=(-1)^{\lambda+j_c+j_\gamma+1} \sqrt{[j_c][j_\alpha][j_\gamma][j_\beta]} 
\,\,\,\xi(l_c +l_\alpha +\lambda )\xi(l_\gamma +l_\beta +\lambda )\nonumber\\
\times \left(	{\lambda \atop 0}{j_c \atop -\frac{1}{2}}
		{j_\alpha \atop \frac{1}{2}}\right) \left( {\lambda \atop 0}
		{j_\gamma \atop -\frac{1}{2} }{j_\beta \atop \frac{1}{2}}\right) 
R_\lambda (c\gamma\alpha\beta )\,
\end{eqnarray}
and
\begin{eqnarray}
\label{eq:radial_integral}
R_\lambda (c\gamma\alpha\beta)
 = \int \int  \frac{r_<^\lambda} {r_>^{\lambda +1}}[f_c (r)f_\alpha (r)+g_c (r)g_\alpha (r)] \nonumber \\
\times  [f_\gamma (r')f_\beta (r')+g_\gamma (r')g_\beta (r')]\rmd r\rmd r'
\end{eqnarray}
is the radial Coulomb integral, $f$ and $g$ being the upper and lower components of the relativistic orbital spinors. $ \alpha\leftrightarrow\beta$ means that the expression inside the square bracket should be evaluated by interchanging the orbitals $\alpha$ and $\beta$ and must be added the former one. The detailed derivation has been presented in our papers \cite{FGGH02,BHDFG15}. The occupation numbers $n_\gamma$, $n_\alpha$ $n_\beta$ are evaluated in the relativistic ground state configuration of the target ion $4d_{3/2}^4 4d_{5/2}^6 4f_{5/2}^{n_1} 4f_{7/2}^{n_2}$.

In \eref{eq:capture-explicit} the summation over doorways $\alpha\beta\gamma^{-1}$ extends over the dielectronic excitations only (in the determinant basis) since the two-body Coulomb matrix elements vanish for the states with more than two excited electrons from the ground configuration $4d^{10}4f^{(n+1)}$. The electron in the continuum with angular quantum numbers $(jl)$ and energy $\varepsilon$ falls into the particle orbital $\alpha$ of the compound state during the capture process. 
The energies of the doorways $\varepsilon _{\alpha\beta\gamma^{-1}}$ are the  configuration average energies given in \eref{eq:CA_energy} with respect to the ionisation threshold: $\varepsilon _{\alpha\beta\gamma^{-1}}=E^\textsl{CA}_{\alpha\beta\gamma^{-1}}-E_i^\textsl{CA}$.
 
\subsection{Fluorescence yield}

The fluorescence yield determines the probability of the radiative stabilization after electron capture to a compound resonance.
It can be obtained from the ratio of the total autoionization width $\Gamma^{(a)}$ and the total radiative width $\Gamma^{(r)}$:
\begin{equation}
\label{eq:yield}
\omega_f=\left(1+\frac{\Gamma^{(a)}}{\Gamma^{(r)}}\right)^{-1}. 
\end{equation}
The statistical theory  for  the electric dipole matrix elements ($E1$ amplitudes) has been developed in Ref. \cite{FGG98} including numerical tests and comparison with the available experimental data.  The captured electron in a compound resonance above the ionisation threshold of the ion has a huge number of  available final  states to make a radiative transition (since the level density is extremely high).  The chaotic (ergodic) mixing between all available many-electron basis states near  given excitation energy guarantees that practically all transitions permitted by the conserving quantum numbers can take place. On the other hand, near electron collision energy $\varepsilon=0$ there is only one autoionization channel, with the final ion in the ground state. This makes the  fluorescence yield   $\omega_f \approx 1$ for $\varepsilon=0$. Increase of $\varepsilon$ leads to opening of new autoionization channels, and $\omega_f $ decreases.

The mean-squared values of the matrix elements of the dipole operator between chaotic many-body states determine the total radiative width which is given by the formula derived in Ref. \cite{FGGH02} (see also \cite{FGG98,FGGK94,GGF99,DFGH12}):
\begin{equation}
\label{eq:Gamma_r}
\Gamma ^{(r)}= \sum _{\alpha ,\beta }
\frac{4\omega _{\beta \alpha }^3} {3c^3}
|\langle \alpha \|d\|\beta \rangle |^2
\overline{\frac{n_\beta }{\left[j_\beta\right]}\left( 1-\frac{n_\alpha }
{\left[j_\alpha\right]}\right)} ,
\end{equation}
The summation here goes over  single-particle orbitals $\beta$ and $\alpha$,  $\left[j_\alpha\right]=2j_\alpha+1$ is the degeneracy of the orbital $\alpha$, $\omega _{\beta \alpha }=\varepsilon _\beta -\varepsilon _\alpha >0$,
$\langle \alpha \|d\|\beta \rangle $ is the reduced matrix element of the dipole operator
between the orbitals $\alpha $ and $\beta $. The line over the expression in \Eref{eq:Gamma_r} means the average of the orbital occupation numbers in the compound states at the electron-ion collision energy $\varepsilon$. 
In our notation the maximal orbital occupation number is $n_\alpha=2j_\alpha+1$. The interpretation of \eref{eq:Gamma_r} is simple: to have a transition the initial orbital $\beta$ should be occupied  and the final orbital $\alpha$ vacant.

The total radiative width \eref{eq:Gamma_r} of the resonances around energy $\varepsilon$ is easy to calculate statistically since it includes matrix elements of a single-particle operator (the electric dipole operator) between orbitals. Therefore the values of the matrix elements are fixed as soon as the DHF orbitals are determined. The statistical nature only comes from the averaging over occupation numbers. The occupation numbers of the orbitals vary with energy, and can be found by  averaging over all compound state components:
$\overline{n}_\alpha(E)=\sum_m \overline{C^2_m}(E)n^{(m)}_\alpha$,
where $n_\alpha^{(m)}$ is the occupation number of the subshell $\alpha$ in the basis
state $m$.  Although there are other ways to find the average of the occupation factor, we have not found a significant differences in the final results obtained by different ways of averaging it.

The total autoionization width of a many-body chaotic state (a compound state)  increases rapidly when the energy of the system increases since a number of low lying levels $i'$  with energy 
$\varepsilon_{i'}<\varepsilon$ in  the target ion $A^{q+}$ 
increases opening new autoionization channels. Together with the width $\Gamma^{(a)}_i$ of the initial channel (corresponding to the decay to the ground state of  the target ion $A^{q+}$), the total autoionization decay width can be  presented as
\begin{equation}
\label{eq:totalautowidth}
\Gamma^{(a)} = \Gamma^{(a)}_i+\sum_{i'(\varepsilon_{i'}\leq\varepsilon)} \Gamma^{(a)}_{i'}.
\end{equation}
Note that $\Gamma_{i'}^{(a)}$ is not a width of a doorway state given in \eref{eq:Gamma_ni}; it is the width of a compound state which can be expressed as the weighted sum of the doorway widths:  $\Gamma_{i'}^{(a)}=\sum_n\overline{|C_n|^2}\Gamma_{ni'}^{(a)}$.
Therefore the autoionization width of  chaotic multi-excited-electron state is suppressed by the  factor $|C_n|^2 \sim  D/\Gamma_{\rm spr} \sim 1/N$, where $N    \sim \Gamma_{\rm spr}/D $ is the  number of the principle components in a compound state.
Our numerical calculation shows that at low energies the total radiative width  of a chaotic compound state is much bigger than the total autoionization width due to  the suppression mentioned above.

To obtain the fluorescence yield we need to calculate the total autoionization width (\ref{eq:totalautowidth}).
A partial autoionization width of a compound state is equal to  the weighted sum of the partial doorway widths: $\Gamma_i^{(a)}=\sum_n\overline{|C_n|^2}\Gamma_{ni}^{(a)}$. 
In \cite{DFGHK13}, the average partial compound state autoionization width was estimated using the formula for the capture cross section \eref{eq:capture-explicit}.  Indeed, we may use the following formula linking the capture cross section and the autoionization width:
\begin{equation}
\label{eq:dccs}
\sigma_c= \frac{2\pi^2}{k^2}\sum_J\frac{(2J+1)}{2(2J_i+1)}
\rho_J\Gamma^{(a)}_{i J} \approx \frac{\pi^2}{k^2(2J_i+1)}\rho\Gamma^{(a)}_i.
\end{equation}
Here $\rho_J$ is the density of states for a fixed angular momentum $J$, $\rho=\sum_J (2J+1) \rho_J$ is the total density of states (it is easier to calculate it in the determinant basis). Thus  we can present the compound state autoionization width for a given autoionization channel $i'$  as
$\Gamma^{(a)}_{i'}\approx (2J_{i'}+1) k^2\sigma_c/ (\pi^2\rho)$. Finally, the total width $\Gamma^{(a)}$ of a resonance around energy $\varepsilon$ is obtained by means of equation \eref{eq:totalautowidth}. 

The contribution to the total autoionization width from each target configuration $i'$ 
is given by  $\Gamma^{(a)}_{i'} = K M_{i'}k^2\sigma_{ci'}/(\pi^2\rho)$, where $M_{i'}$ is the number of states in a given configuration with the energy $\varepsilon_{i'}<\varepsilon$. By definition $M_{i'} \leq N_{i'}$, where $N_{i'}$ is the total number of states in this configuration. Here $\sigma_{ci'}$ is the capture cross section for the configuration $i'$ of the target ion. The cross section $\sigma_{ci'} $ has been calculated using \eref{eq:capture-explicit} by constructing the dielectronic doorway states $\alpha\beta\gamma^{-1}$  (i.e, states with two electrons and one hole orbitals) from the target configuration $i'$ and the continuum electron. 
Our approximate \emph{ab initio} calculation gives the coefficient $K=1$. We have attempted to refine the value of $K$  using available experimental data, obtaining $K = 1/2$ for W$^{18+}$ and $K=1$ for W$^{20+}$. However, we could not get any value of $K$ for W$^{19+}$ due to a difference in the energy dependence of the measured and calculated rate. Therefore, we take the coefficient  $K=1$ (\emph{ab initio} value) for W$^{19+}$ and other ions as our prediction.

\begin{figure}[t!]
\centering
\includegraphics[width=0.9\textwidth]{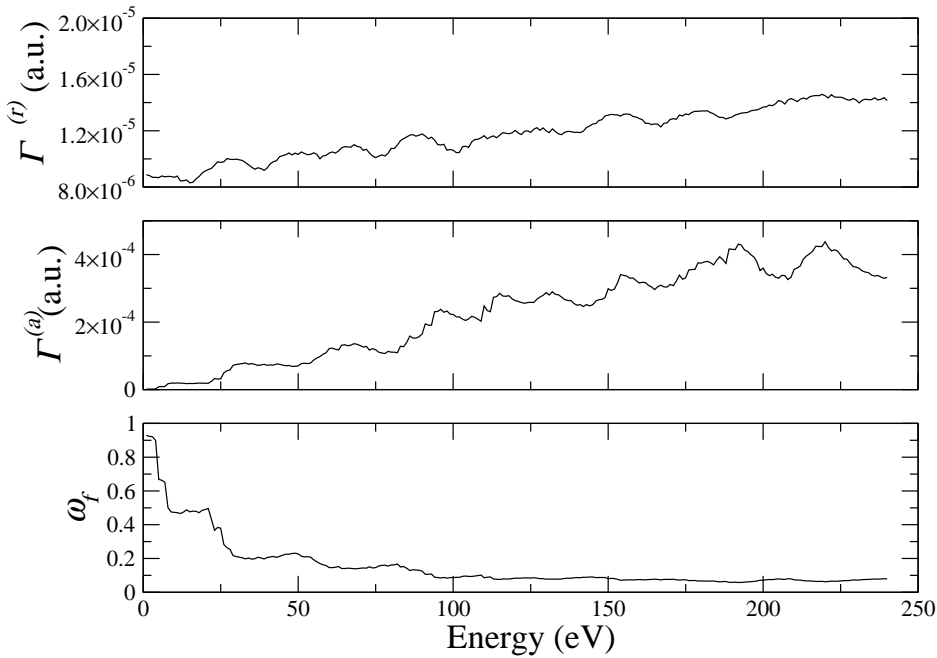}
\caption{Total radiative width $\Gamma^{(r)}$, total autoionization width $\Gamma^{(a)}$, and fluorescence yield $\omega_f$ for electron recombination with W$^{18+}$ ion.}
\label{fig:yield}
\end{figure}

As an example, the total radiative width, the total autoionisation width, and the fluorescence yield $\omega_f$ are plotted in \Fig{fig:yield} for W$^{18+}$. The electron recombination cross section $\sigma(\varepsilon)$ for an ion at energy $\varepsilon$ is calculated by multiplying the capture cross section $\sigma_c(\varepsilon)$ by the fluorescence yield $\omega_f(\varepsilon)$. 

\section{Results and discussion}
\label{sec:results}
We have applied our statistical theory to calculate recombination rate coefficients of the collision of an electron with tungsten ions with open $f$ shells. Experimental measurements for the three ions W$^{20+}$, W$^{18+}$, and W$^{19+}$ have been published in references \cite{SBM11,SBKN14,BSKN16}. The calculations of other groups for these ions have been reported in \cite{BBGO12,SBKN14,BSKN16}. We have obtained the results for these and five other tungsten ions in the present work.

The calculated recombination cross section has to be thermally averaged  to compare with the measured data \cite{SMGL01}: $\alpha(v_{\rm rel})=\int v\sigma(v)f(v_{\rm rel}, {\bf v})d^3v$, where $v_{\rm rel}$ is the average longitudinal velocity of the electron beam relative to the ions. The electron velocity distribution $f(v_{\rm rel}, {\bf v})$ depends on the temperature of the beam. If the transverse temperature $T_\bot$ is much higher than the longitudinal one $T_\|$ (e.g., $k_BT_\bot=10$ meV and $k_BT_\|=0.15$ meV for the experiment~\cite{SBM11}) and $v_{\rm rel}^2\gg k_BT_\bot$, we may assume that the electron recombination rate can be obtained as $\alpha\approx v_{\rm rel} \sigma=\sigma\sqrt{2\varepsilon/m}$ \cite{FGGH02}.
 
\begin{figure}[t!]
\centering
\includegraphics*[width=0.9\textwidth]{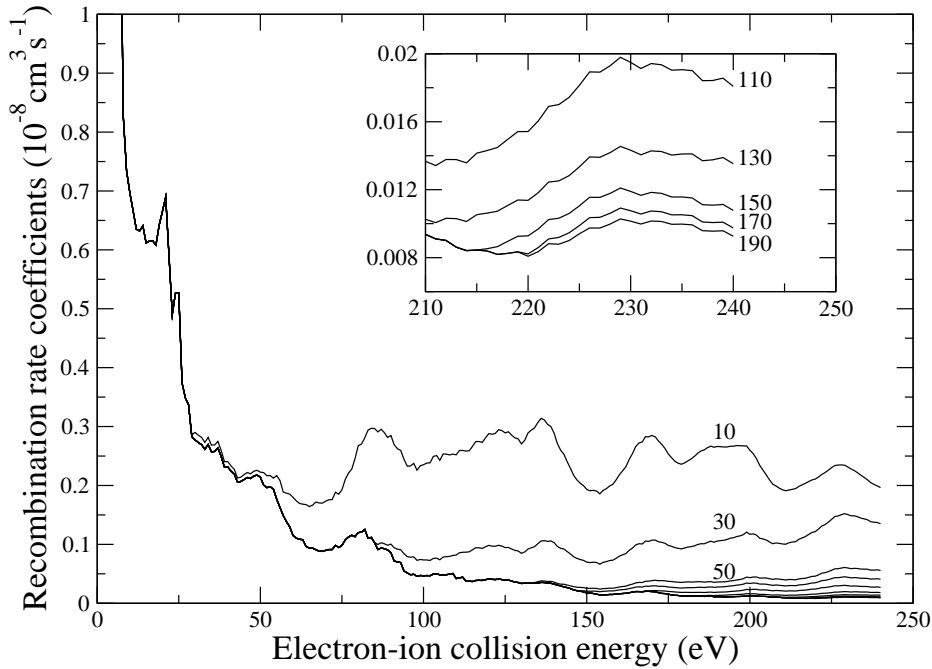}
\caption{Convergence of the recombination rate coefficient with respect to the number of configurations providing the autoionization channels for W$^{18+}$ in the calculation energy range (240 eV).  The inset shows the magnified tail part of the curves around 240 eV. The numbers associated with the curves are the numbers of configurations of the target ions used to calculate the total autoionization rate $\Gamma^{(a)}$.}
\label{fig:channel}
\end{figure}

In \Fig{fig:channel} we present the convergence of $\alpha$ with respect to the number of configurations of the target ion available in 240 eV energy range. We see that low energy configurations give the dominant contribution to the recombination rate since higher configurations produce a small number of open autoionization channels. The inset of the figure shows that the calculation of $\alpha$ with 150 configurations practically saturates the result even though there are 227 target configurations giving contribution to the autoionization in the energy range of 240 eV above the ionisation threshold.

\begin{figure}[t!]
\centering
\includegraphics*[width=0.9\textwidth]{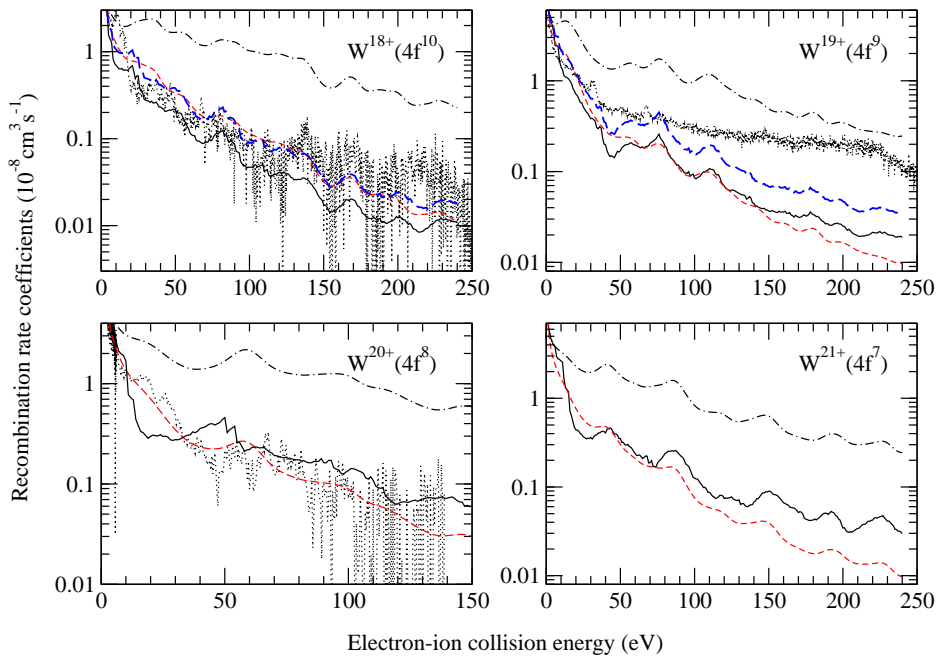}
\caption{(Color online) The calculated electron recombination rate (black solid) obtained from $\alpha=\omega_f\alpha_c$. The capture rate (black dot-dashed) has been calculated using \Eref{eq:capture-explicit} as $\alpha_c=\sigma_c\sqrt{2\varepsilon/m}$. The fluorescent yield $\omega_f$ is given by \eref{eq:yield}. Available experimental data for three ions are shown by doted lines. The blue long-dashed lines are obtained by replacement of  the total autoionization width $\Gamma^{(a)}$ with  $\Gamma^{(a)}/2$ in \eref{eq:yield} for the fluorescent yield to get a better agreement with the experimental measurements for W$^{18+}$ (see text). The red short-dashed lines are obtained using a simple expression for the fluorescence yield $\omega_f(\varepsilon)=1/(1+b \varepsilon)$~\cite{BHDFG15} which fits the experimental recombination rate data. Here $b=0.08$~eV$^{-1}$ for W$^{18+}$, $0.1$~eV$^{-1}$ for W$^{19+}$, $0.124$~eV$^{-1}$ for W$^{20+}$, and $0.1$~eV$^{-1}$ for W$^{21+}$.}
\label{fig:rate1}
\end{figure}

\begin{figure}[t!]
\centering
\includegraphics*[width=0.9\textwidth]{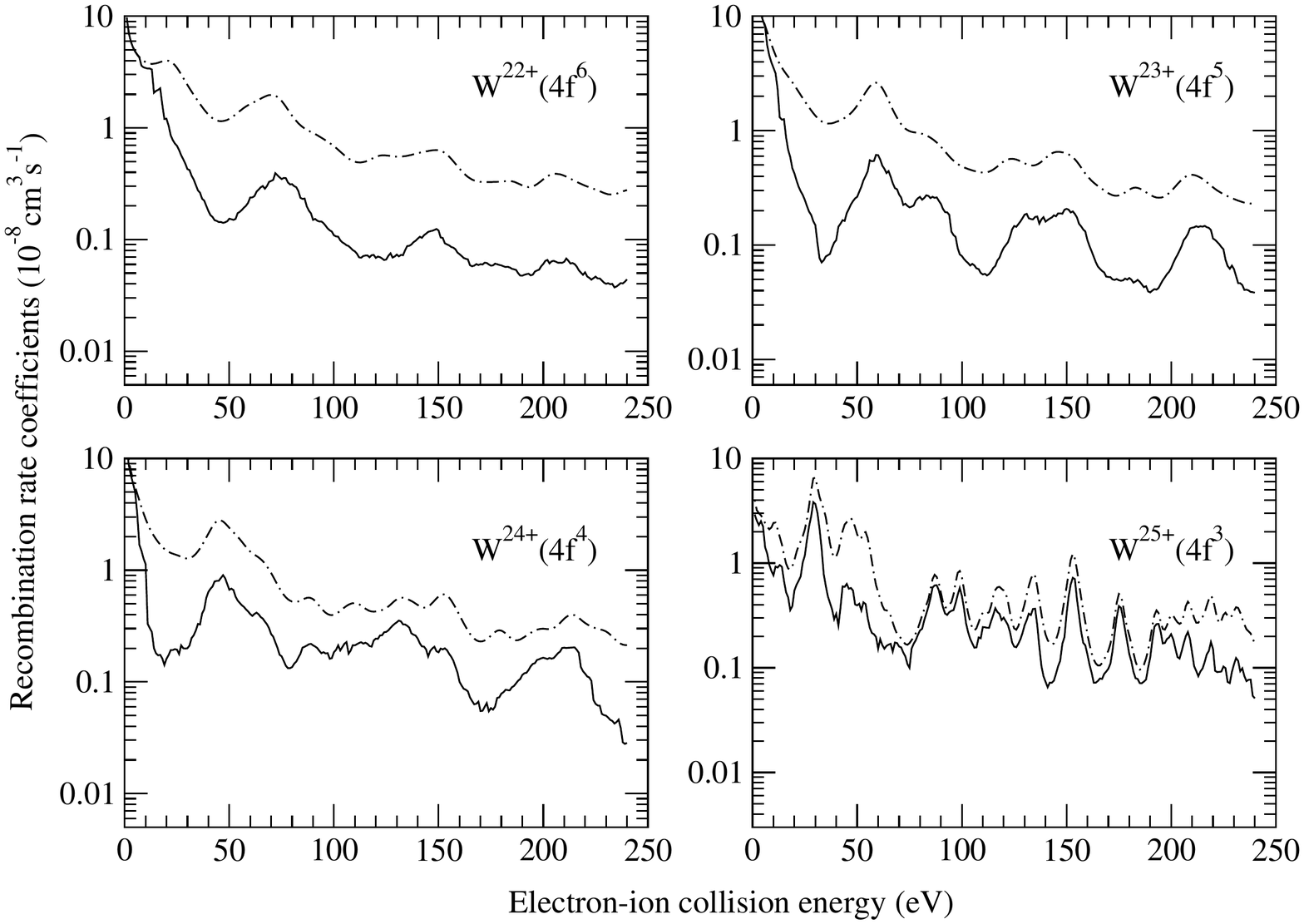}
\caption{The predicted electron recombination rate coefficients (solid lines). The dot-dashed curves show the electron capture rates (the difference between these two curves is due to the fluorescent yield $\omega_f<1$).}
\label{fig:rate2}
\end{figure}

In \Fig{fig:rate1} our calculated electron recombination coefficients are compared with experimental measurements for W$^{18+}$, W$^{19+}$, and W$^{20+}$ (results for W$^{21+}$ are also included here). \Fig{fig:rate2} shows our predictions for the recombination rates for the ions W$^{22+}$, W$^{23+}$, W$^{24+}$, and W$^{25+}$ where there are no experimental data. Our final results for the electron recombination rates (black solid lines) are in good agreement with the experiment for W$^{18+}$ and W$^{20+}$. However, the energy dependence of the calculated rate for  W$^{19+}$ is different from the measurement. Surprisingly, we see that the energy dependence of the experimental data for W$^{19+}$ is also different from those for W$^{18+}$ and W$^{20+}$.

\fulltable{\label{t:fit}The fitting parameters of the calculated plasma rate coefficients in the expression Eq.(\ref{eq:fit})} 
\br
Target ion & \centre{6}{${\rm cm}^3{\rm s}^{-1}({\rm eV})^{3/2}\times 10^{-7}$} & \centre{6}{(eV)} \\
\ns
&\crule{6}&\crule{6}\\
W$^{q+}$ & $C_1$ & $C_2$ & $C_3$ & $C_4$ & $C_5$ & $C_6$ & $E_1$ & $E_2$ & $E_3$ & $E_4$ & $E_5$ & $E_6$ \\
\mr 
W$^{18+}$  & 1.4119 & 4.0146 & 7.6264 & 10.3072 & 9.4524 & 3.9021 & 0.9794 & 5.5671 & 18.4505 & 44.2112 & 96.8032 & 192.2218 \\
W$^{19+}$  & 1.2532 & 4.8403 & 6.2878 &  6.7555 & 9.7919 & 4.6861 & 1.0277 & 4.9220 & 14.3877 & 36.8638 & 88.9257 & 188.0938 \\
W$^{20+}$  & 1.1491 & 4.5797 & 5.0636 &  6.4154 & 9.5361 & 4.8678 & 0.2413 & 2.9154 &   9.5268 & 33.5650 & 69.5228 & 119.9227 \\            
W$^{21+}$  & 1.5591 & 4.9246 & 7.5387 & 9.4209 & 12.9804 & 7.7875 & 1.2178 & 4.4254 & 11.9336 & 37.8250 & 88.7261 & 190.5227 \\          
W$^{22+}$  & 2.6048 & 5.9589&10.1782 & 9.3156 &14.6874 & 10.6896 & 1.4464 & 5.1320 & 11.4593 & 25.4832 & 79.3531 &185.3436 \\           
W$^{23+}$  &2.9095 & 8.5022 & 11.7326 & 6.4043 & 17.7008 & 14.7894 & 1.0545 & 3.3416 & 8.9154 & 31.1050 & 84.0439 & 190.5443\\       
W$^{24+}$  &1.9613 & 6.6207 & 5.0859 & 8.7325 & 20.0231 & 21.9684 & 1.0384 & 4.0949 & 21.2570 & 49.1594 & 86.6538 & 186.2551\\               
W$^{25+}$  &1.1874& 3.3312& 15.0977&12.9584& 25.5774 & 20.5568 & 1.7333 & 6.6740 & 44.5367 & 23.4193 & 109.7857 &197.1963\\             
\br
\endfulltable

According to a private communication of S. Schippers the recombination data for the electron energy above 50 eV have a significant  uncertainty due to  large statistical errors (due to a small number of events) and a background from other processes which should be subtracted. For example, this resulted in a significant difference between the preliminary experimental data for the $W^{19+}$ recombination \cite{S15} and the final data published in Ref. \cite{BSKN16} and presented on Fig. 4. In this situation the theoretical calculation may be an important source of the recombination data for higher energies.

We have proposed a simple energy dependence of the fluorescence yield in our previous work as $\omega_f=1/(1+b\varepsilon)$. It has been shown that a single constant $b$ is sufficient to fit the experimental curve for the capture rate. We found  $b=0.124$~eV$^{-1}$ for W$^{20+}$ \cite{BHDFG15}. In the present work we  have found  $b=0.08$~eV$^{-1}$ for W$^{18+}$. These fits and our calculations have the same energy dependence. For W$^{21+}$ there are no published experimental data and we use  $b=0.1$ (eV)$^{-1}$. We have also taken the average value $b=0.1$ for W$^{19+}$ since this seems  natural for the ion between W$^{18+}$ and W$^{20+}$. Note that all rates obtained using these fits (red short-dashed lines in \Fig{fig:rate1}) are in agreement with our calculated rates (black solid lines).
 
\begin{figure*}[t!]
\centering
\includegraphics*[width=0.9\textwidth]{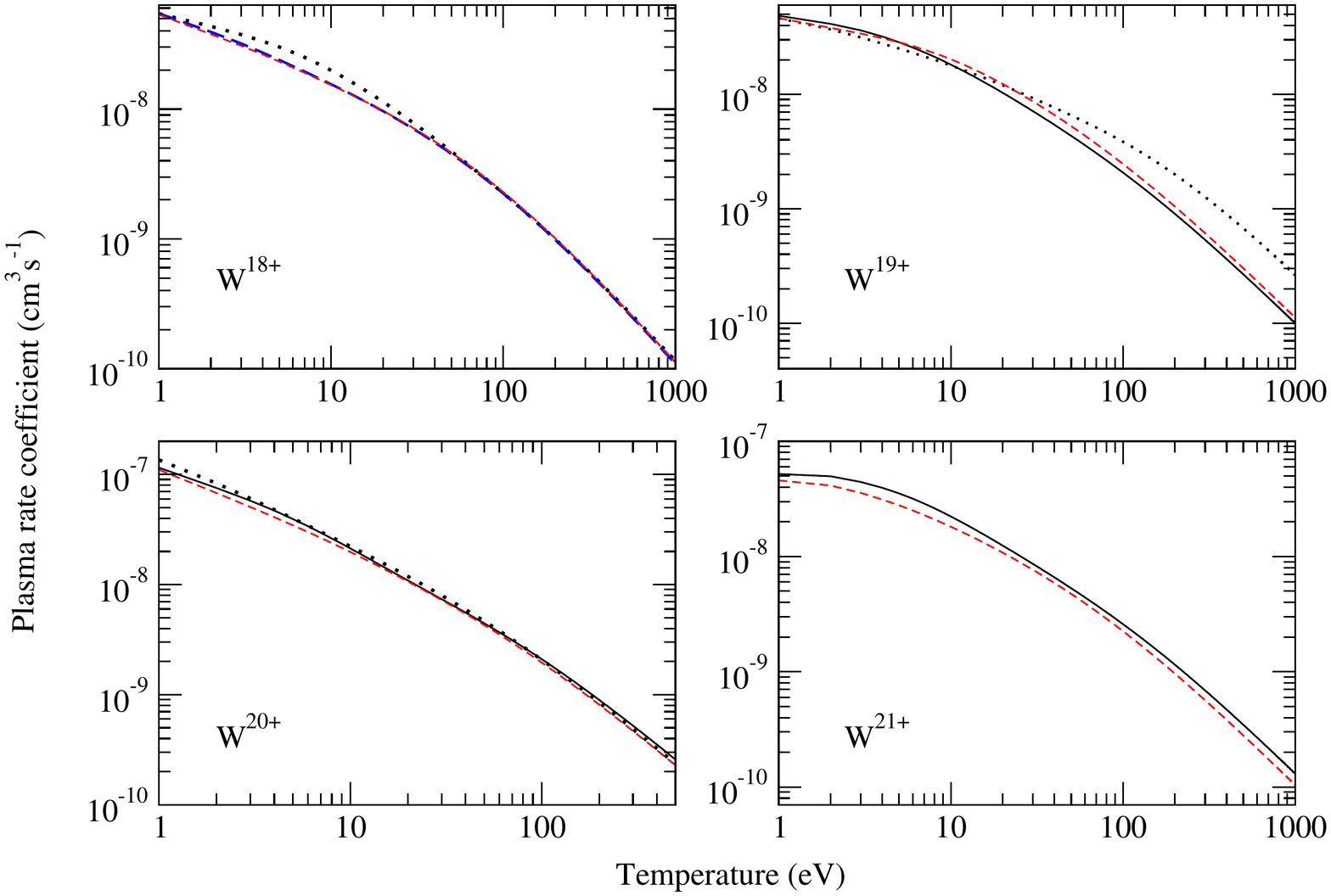}
\caption{(Color online) Experimental (black doted-line) and calculated plasma recombination rates (black solid lines). The calculated rate for W$^{18+}$ are shown with blue long-dashed line since the coefficient $K=1/2$ instead of 1 for others.
The red short-dashed lines are obtained using  a simple expression for  the fluorescence yield $\omega_f(\varepsilon)=1/(1+b \varepsilon)$ with  $b=0.08$ (eV)$^{-1}$ for W$^{18+}$, $0.1$ (eV)$^{-1}$ for W$^{19+}$, $0.124$ (eV)$^{-1}$ for W$^{20+}$, and $0.1$ (eV)$^{-1}$ for W$^{21+}$ (see  Fig.\ref{fig:rate1})}
\label{fig:temrate1}
\end{figure*}

\begin{figure*}[t!]
\centering
\includegraphics*[width=0.9\textwidth]{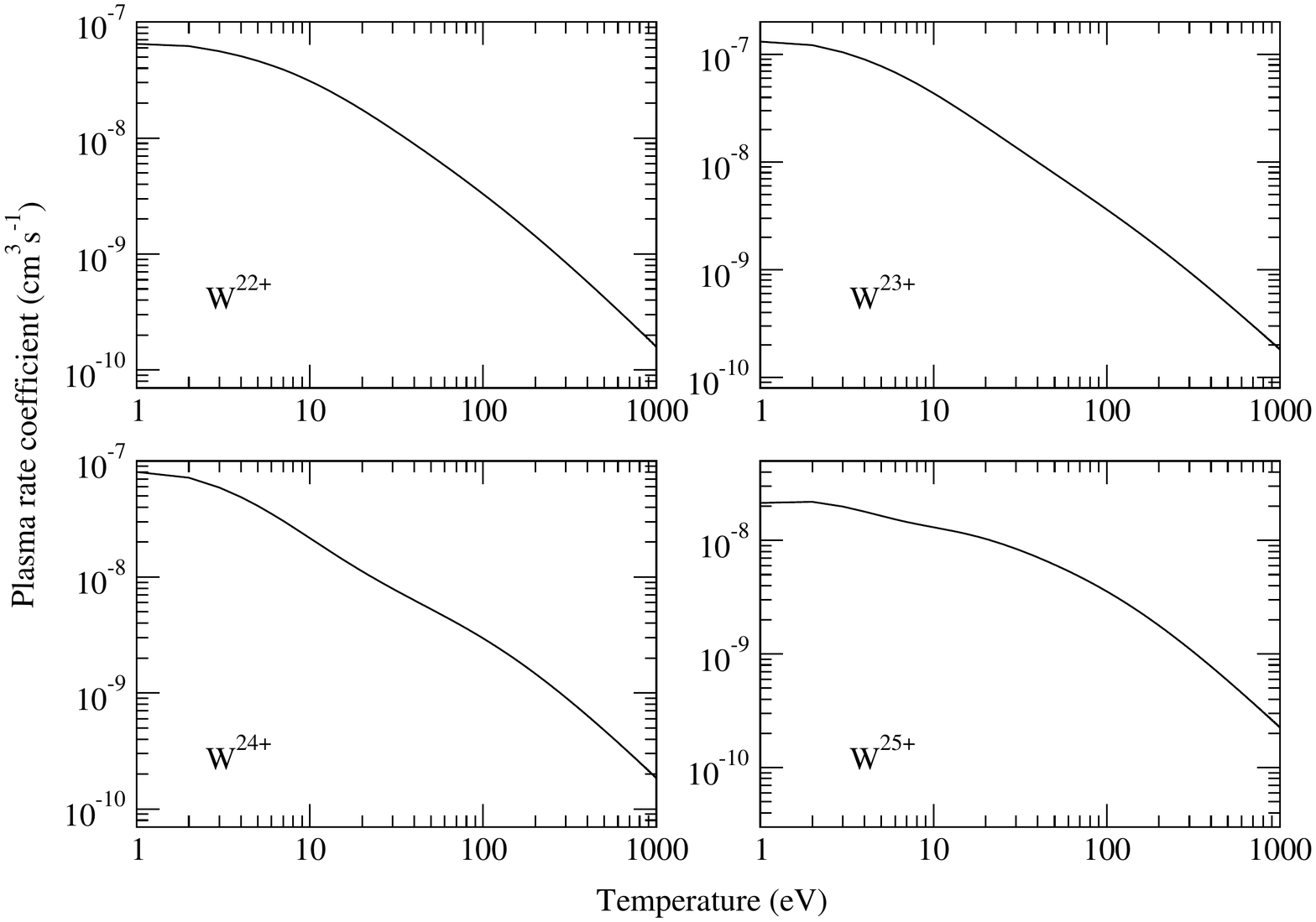}
\caption{The plasma rate coefficients corresponding to the merged-beam recombination rates plotted on Fig. \ref{fig:rate2}. }
\label{fig:temrate2}
\end{figure*}

Finally, the plasma rate coefficient is calculated by convolution of the recombination cross section with an isotropic Maxwellian electron energy distribution since the recombination process takes place in hot plasma \cite{SMGL01}. It can be calculated by taking integration of merged-beam electron recombination rate $\alpha(\varepsilon)$ as 
\begin{equation}
\label{eq:plasma}
\alpha(T)=\frac{2(k_BT)^{-\frac{3}{2}}}{\sqrt{\pi}}\int_0^\infty d\varepsilon~\alpha(\varepsilon)\sqrt{\varepsilon}~\exp\left(-\frac{\varepsilon}{k_BT}\right),
\end{equation} 
where $T$ is the plasma electron temperature and $k_B$ is the Boltzmann constant. In this work we have performed the calculation starting from 1 eV to 240 eV in 1~eV energy steps. In one of our previous works~\cite{DFGH12} we have shown that for small energies (less than 1 eV) our theory recovers the enhancement in the measured recombination rate. Therefore, we  have extrapolated our result for the rate to the energy interval of $0<\varepsilon\le 1$ eV without calculation, just to estimate the contribution of energies below 1 eV to the integral. 
We have not done this extrapolation for other ions instead they are taken constant in less than 1 eV. This is acceptable since a significant abundance of highly charged tungsten ions in plasma appears at temperatures above 100 eV, where the energy integration interval of $0<\varepsilon\le 1$ eV is not important.
The results are presented at Figs.~\ref{fig:temrate1} and \ref{fig:temrate2}. 

We fitted our calculated plasma rate coefficients $\alpha(T)$ with the standard formula \cite{SBKN14,SMGL01} to make them easy to use in the modelling of the fusion plasma. Twelve parameters $C_i$ and $E_i$, $(i=1,...,6)$ are used in the exponential sum
\begin{equation}
\label{eq:fit}
\alpha(T)=(k_BT)^{-\frac{3}{2}}\sum_{i=1}^{6}C_i~\exp\left(-\frac{E_i}{k_BT}\right),
\end{equation} 
to fit our results perfectly in the temperature interval 1-1000 eV. They are tabulated in Table \ref{t:fit} for each tungsten ion considered in the current work.

\section{Conclusion}
\label{sec:conc}
The present paper provides the first many-body calculations of the electron recombination rates for the highly charged tungsten ions W$^{q+}$, $q = 21$ -- $25$ with open f-shells. The experimental data for these rates are not available and it is difficult to measure them using existing experimental techniques. Furthermore at electron energy above 50 eV, the experimental rates have a significant uncertainty due to large statistical errors and difficult subtraction of background. Therefore the current calculations are needed for modelling of the plasma in thermonuclear reactors, which is always contaminated by tungsten ions. 

We explore a mechanism of multi-electron recombination (extension of the di-electronic recombination) due to a very dense spectrum of chaotic compound resonances. We have developed a statistical theory to describe such processes. Our results are presented in Figs. \ref{fig:rate1} -- \ref{fig:temrate2}. We fit the calculated  plasma rate coefficients by the standard analytical formulae presented in Eq. \ref{eq:fit} to make the use of our results simple and convenient. The fitting parameters $C_i$ and $E_i$, $(i=1,...,6)$ of this formula for each tungsten ion are given in the Table \ref{t:fit}.

To test our theory we compare our results with the available  experimental data for  W$^{18+}$, W$^{19+}$, and W$^{20+}$. We show that our statistical theory provides a reasonable quantitative description of the recombination rates for ions with open f-shell which possess chaotic many-electron compound resonances. This multi-electron recombination rate exceeds the direct recombination rate by 2-3 orders of magnitude.
\newpage
\ack
This work was funded in part by the Australian Research Council and the Gutenberg Fellowship. We thank S. Schippers for providing experimental data in numerical form and to M. Kozlov for useful comments.


\section*{References}
\begin{thebibliography}{99}

\bibitem{ADAS} Atomic Data and Analysis Structure (ADAS), http://www.adas.ac.uk/

\bibitem{BOSA11} Badnell N R, O'Mullane M G, Summers H P, Altun Z, Bautista M A, Colgan J, Gorczyca T W, Mitnik D M, Pindzola M S and Zatsarinny O 2003 {\it Astron. Astrophys.} {\bf 406} 1151 

\bibitem{PCEHK13} Pitts R A, Carpentier S, Escourbiac F, Hirai T, Komarov V, Lisgo S, Kukushkin A S, Loarte A, Merola M, Naik A S, Mitteau R, Sugihara M, Bazylev B and Stangeby P C 2013 {\it J. Nucl. Mater.} {\bf 438} S48

\bibitem{RJC15} Romanelli F on behalf of JET Contributors 2015 {\it Nucl. Fusion} {\bf 55} 104001

\bibitem{BHDFG15} Berengut J C, Harabati C, Dzuba V A, Flambaum V V and Gribakin G F 2015  
{\it Phys. Rev. A} {\bf 92} 062717

\bibitem{FGGK94} Flambaum V V, Gribakina A A, Gribakin G F and Kozlov M G 1994 {\it Phys. Rev. A} {\bf 50} 267

\bibitem{GGF99} Gribakin G F, Gribakina A A and Flambaum V V
1999 {\it Aust. J. Phys.} {\bf 52} 443 (arXiv:physics/9811010)

\bibitem{DFGH12} Dzuba V A, Flambaum V V, Gribakin G F, Harabati C, 
2012 {\it Phys. Rev. A} {\bf 86} 022714

\bibitem{FKG15} Flambaum V V, Kozlov M G and Gribakin F G 2015 {\it Phys. Rev. A} {\bf 91} 052704

\bibitem{FGGH02} Flambaum V V, Gribakina A A, Gribakin G F and Harabati C 2002 {\it Phys. Rev. A} {\bf 66} 012713

\bibitem{HUSF98} Hoffknecht A, Uwira O, Schennach S, Frank A, Haselbauer J, Spies W, Angert N, Mokler P H, Becker R, Kleinod M, Schippers S and M\"uller A 1998 {\it J. Phys. B} {\bf 31} 2415

\bibitem{SBM11} Schippers S, Bernhardt D, M\"uller A, Krantz C,
Grieser M, Repnow R, Wolf A, Lestinsky M, Hahn M, Novotn\'y O and Savin D W 2011 {\it Phys. Rev. A} {\bf 83} 012711

\bibitem{DFGHK13} Dzuba V A, Flambaum V V, Gribakin G F, Harabati C and Kozlov M G 2013 
{\it Phys. Rev. A} {\bf 88} 062713

\bibitem{BBGO12} Badnell N R, Ballance C P, Griffin D C and O'Mullane M 2012 {\it Phys. Rev. A} {\bf 85} 052716

\bibitem{SBKN14} Spruck K, Badnell N R, Krantz C, Novotn\'y O, Becker A, Bernhardt D, Grieser M, Hahn M, Repnow R, Savin D W, Wolf A, M\"uller A and Schippers S 2014 {\it Phys. Rev. A} {\bf 90} 032715

\bibitem{BSKN16} Badnell N R, Spruck K, Krantz C, Novotn\'y O, Becker A, Bernhardt D, Grieser M, Hahn M, Repnow R, Savin D W, Wolf A, M\"uller A and Schippers S 2016 {\it Phys. Rev. A} {\bf 93} 052703

\bibitem{S15} Spruck K 2015 {\it Dielectronic Recombination Experiments with Tungsten Ions at the Test Storage Ring and Development of a Single-Particle Detector at the Cryogenic Storage Ring} (PhD Thesis)

\bibitem{ZBFH96} Zelevinsky V, Brown B A, Frazier M and Horoi M 1996
{\it Phys. Rep.} {\bf 276} 85 

\bibitem{BM69} Bohr A and Mottelson B 1969 {\it Nuclear structure} Vol. 1 (New York: Benjamin)

\bibitem{F93} Flambaum V V1993 {\it Physica Scripta T: Proceedings of the 1992 Nobel Symposium} {\bf 46} 198

\bibitem{FV93} Flambaum V V and Vorov O K 1993 {\it Phys. Rev. Lett.} {\bf 70} 4051

\bibitem{FG95} Flambaum V V and Gribakin G F 1995 {\it Progress in 
Particle and Nuclear Physics} {\bf 35} 423

\bibitem{FG00} Flambaum V V and Gribakin G F 2000
{\it Philos. Mag. B} {\bf 80} 2143 

\bibitem{GS03} Gribakin G F and Sahoo S 2003 {\it J. Phys. B} {\bf36} 3349

\bibitem{NIST} Kramida A, Ralchenko Yu, Reader J and NIST ASD Team 2015 {\it NIST Atomic Spectra Database}
(ver. 5.3), [Online]. Available:
{\tt{http://physics.nist.gov/asd}} [2016, May 18].
National Institute of Standards and Technology
Gaithersburg, MD

\bibitem{MB42} Massey H S W and Bates D R 1942 {\it Rep. Prog. Phys.} {\bf 9} 62

\bibitem{LL77} Landau L D and Lifshitz E M 1977 {\it Quantum Mechanics} 3rd ed.
(Oxford: Pergamon)

\bibitem{FGG98} Flambaum V V, Gribakina A A and Gribakin G F 1998 {\it Phys. Rev. A} {\bf 58} 230

\bibitem{SMGL01} Schippers S, M\"uller A, Gwinner G, Linkemann J, Saghiri A A and Wolf A 2001 {\it Astrophys. J.} {\bf 555} 1027

\end{thebibliography}
\end{document}